\documentclass[a4paper,twocolumn]{article}

\usepackage{epsfig}
\usepackage{amssymb}

\begin{document}

\title{Control of neural chaos by synaptic noise}
\author{J. M. Cortes$^{\dag \ddag}$, J. Marro$^{\dag}$ and J. J. Torres$^{\dag}$ \\
$^{\dag}$Institute \textit{Carlos I} for Theoretical and Computational Physics, and \\
 Departamento de Electromagnetismo y F\'{\i}sica de la Materia,\\ 
University of Granada, E-18071 Granada, Spain.\\
$^{\ddag}$Department of Biophysics,  Radboud University of Nijmegen, \\
6525 EZ Nijmegen, The Netherlands}

\maketitle

{To appear in Biosystems, 2005}

{Corresponding author: Jesus M. Cortes}

{mailto:jcortes@ugr.es}

\begin{abstract}

We studied \textit{neural automata} ---or neurobiologically inspired
cellular automata--- which exhibits chaotic itinerancy among the different
stored patterns or memories. This is a consequence of activity-dependent
synaptic fluctuations, which continuously destabilize the attractor and
induce irregular hopping to other possible attractors. The nature of the
resulting irregularity depends on the dynamic details, namely, on the
intensity of the synaptic \textquotedblleft noise\textquotedblright\ and on
the number of sites of the network that are synchronously updated at each
time step. Varying these details, different regimes occur from regular to
chaotic. In the absence of external agents, the chaotic behavior may turn
regular after tuning the noise intensity. It is argued that a similar
mechanism might be at the origin of the self-control of chaos in natural
systems.

\end{abstract}
\section{The model and its motivation}

We report on the complex dynamics and possible applications of a novel 
\textit{neural automaton} or cellular automaton \cite{wolframNATURE}
inspired in neurobiology. The model exhibits dynamic associative
memory, including kind of switching behavior that has been reported for
neural networks with dynamic synapses \cite{torresNC}, \cite{cortesNEUCOM}, 
\cite{abbottNATURE}. Our automaton incorporates fast fluctuations of
synaptic intensities which depend on neuron activity. Such \textquotedblleft
noise\textquotedblright\ induces instability of the recalling dynamics in a
way that mimics how the brain efficiently solves some complex tasks. In
fact, a rapid response to highly changing stimuli is believed to play a
functional role during both attention and sequential processing of parallel
sensory information \cite{cortesNC}. In this report, we adapt a previous
proposal \cite{marroPRE} to show that fast synaptic noise can control the
complexity and chaoticity of dynamics and, in particular, the details of the
temporal oscillations of the neural activity. Unlike in earlier work \cite%
{molgedeyPRL}, \cite{schiffNATURE}, \cite{freemanIEEE} the noise intensity
in the present mechanism varies autonomously, which could be more relevant
to the self-control of chaos in neural systems as well as in other cases.

The model consists of $N$ cooperative and, for simplicity, fully--connected
neurons with stochastic dynamics\footnote{%
Some consequences of other network topologies have been studied in \cite%
{torresNEUCOM}, for instance.}. A main feature is that, at each time step $t$%
, the individual states of $n\leq N$ neurons are simultaneously updated.
This is performed according to a modification of the Hopfield prescription 
\cite{amariIEEE}, \cite{hopfieldPNAS}, \cite{amitB}. We assume that each
neuron $s_{i},$ endures a current or a local field \cite{gardinerB}, \cite%
{bibitchkov}: 
\begin{equation}
\overline{h_{i}}(\mathbf{S})\equiv \int_{\mathbf{X}}h_{i}(\mathbf{S},\mathbf{%
X})\tilde{P}(\mathbf{X}|\mathbf{S})\mathbf{\mathrm{d}}\mathbf{X.}
\label{efflf}
\end{equation}%
Here, $\mathbf{S}=\left\{ s_{i};i=1,...,N\right\} $ is a neuron
configuration and $\mathbf{X}=\left\{ x_{i}\right\} $ stands for a set of
random variables, $\mathbf{X}=\left\{ x_{i}\right\} ,$ each affecting a
postsynaptic neuron, of distribution $\tilde{P}(\mathbf{X}|\mathbf{S}).$
This amounts to assume short--time, rapid synaptic fluctuations which, in
fact, are known to influence and often determine the neuron activity in many
natural processes. See \cite{marroB} for a technical justification of (\ref%
{efflf}), and \cite{abbottNATURE} for a recent discussion on the role of
synaptic noise, for instance.

This model has already been analyzed both analytically and numerically for
certain choices of parameters. In particular, the case $n=1$ of
\textquotedblleft sequential updating\textquotedblright\ was shown to
exhibit complex hopping between the attractors in some cases \cite{cortesNC}
, and we recently demonstrated \cite{marroPRE} that the hopping may become
chaotic for Little dynamics, namely, $n=N.$ We here illustrate a typical
situation between these two limits by means of computer simulations. The
case with $1<n<N$ for which we present some results here happens to be
relevant to understand the possibility of controlling chaos of the neural
activity by means of synaptic \textquotedblleft noise\textquotedblright .

In order to deal with model simulations that remain versatile enough, we
need to introduce some simplifications in the following; notice, however,
that some of them may turn irrelevant to the resulting emergent behavior.
Most convenient is to restrict ourselves to binary neurons, i.e., $s_{i}=\pm
1,$ which are known to capture the essentials of cooperative phenomena \cite%
{abbott2states}, \cite{torresNC}. Concerning the stochastic variable, we
need to determine both its nature and its distribution. A simple choice is
to assume that synaptic intensities are of the form $w_{ij}=w_{ij}^{\mathrm{L%
} }x_{j}\,$where $w_{ij}^{\mathrm{L}}$ are average weights which, also for
the sake of simplicity, we shall consider to be of the Hebbian type. That
is, $w_{ij}^{\mathrm{L}}=N^{-1}\sum_{\mu }\xi _{i}^{\mu }\xi _{j}^{\mu },$
where $\xi _{i}^{\mu }$ (with $\mu =1,...,M)$ stands for $M$ (binary)
patterns that are assumed hereafter to be \textit{\ stored} in the system.
It then naturally follows stochasticity of the presynaptic currents in (\ref%
{efflf}) which are given by $h_{i}(\mathbf{S},\mathbf{X})=\sum_{j\neq
i}w_{ij}^{ \mathrm{L}}x_{j}s_{j}.$ This is consistent with actual features
of natural systems such as, for example, variations of the glutamate
concentration in the synaptic cleft, and differences in the potency released
from different locations on the active zone of the synapses \cite%
{franksJNEUROSCI}. These and similar \textquotedblleft
noises\textquotedblright\ which cause synaptic fluctuations are typically
very fast compared to the time relaxation of the whole neuron system.
Therefore, it seems sensible to assume that, in the time scale for the
neuron activity, neurons behave as in the presence of a steady distribution
for the synaptic fluctuations. This is taken into account by means of the
distribution $\tilde{P}(\mathbf{X}|\mathbf{S})$ in (\ref{efflf}), a
situation which is discussed with further detail in \cite{marroB}.

\section{Synaptic noise}

Recent neurobiological findings \cite{abbottNATURE}, concerning
activity-dependent processes may help in determining $\tilde{P}(\mathbf{X}|%
\mathbf{S}).$ In particular, it was reported short-time synaptic \textit{%
depression} \cite{tsodyksNC}, i.e., that synaptic weights tend to decrease
under repeated presynaptic activation. A simple way of implementing this in (%
\ref{efflf}) is by taking 
\begin{equation}
\tilde{P}(\mathbf{X}|\mathbf{S})=\prod_{i}\left\{ p\left( \vec{\mathbf{m}}%
\right) \mathrm{\ }\delta (x_{j}+\Phi )+\left[ 1-p\left( \vec{\mathbf{m}}%
\right) \right] \mathrm{\ }\delta (x_{j}-1)\right\} ,  \label{bimod}
\end{equation}%
where the factorization is for simplicity and $\vec{\mathbf{m}}=\vec{\mathbf{%
\ m}}(\mathbf{S})$ is the $M$-dimensional overlap vector of components $%
m^{\mu }(\mathbf{S})=N^{-1}\sum_{i}\xi _{i}^{\mu }s_{i}.$ In accordance with
the mentioned observation, (\ref{bimod}) implies that increasing the mean
firing rate, which will increase the probability function $p\left( \vec{%
\mathbf{m}}\right) ,$ will make more likely that synaptic intensities
decrease by a factor of $\Phi .$ The Hopfield model, for which such
depressing noise is absent, corresponds here to the limit $\Phi \rightarrow
-1$. Finally, in order to fully determine the model, one may use the choice 
\cite{cortesNC} $\zeta \left( \vec{\mathbf{m}}\right) =\left( 1+\alpha
\right) ^{-1}\sum_{\nu }\left[ m^{\nu }\left( \mathbf{S}\right) \right]
^{2}, $ where $\alpha =M/N$ is the network load parameter \cite{hertzB}.
After some straightforward algebra, one obtains the effective currents as 
\begin{equation}
\overline{h_{i}}(\mathbf{S})=\left( 1-\frac{1+\Phi }{1+\alpha }\sum_{\mu }%
\left[ m^{\mu }\left( \mathbf{S}\right) \right] ^{2}\right) \sum_{\mu }\xi
_{i}^{\mu }m^{\mu }\left( \mathbf{S}\right) .  \label{lfaprox}
\end{equation}

\begin{figure}[tbp]
\centerline{
\psfig{file=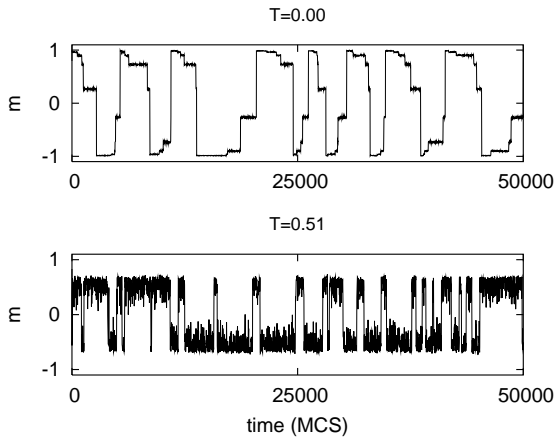,width=7.5cm}}
\caption{{\protect\small This shows the time variation of the overlap $%
m\equiv m^{1}(\mathbf{S})$ between the current neural activity, $\mathbf{S,}$
and the only pattern which is \textit{stored }in the synaptic weights, i.e.,
for $M=1$, as obtained in a Monte Carlo simulation with $N=3600$ neurons and
a depressing factor $\Phi =0.043.$ The top graph is for $T=0,$ i.e., in the
absence of thermal fluctuations, while the bottom graph is for $T=0.51.$}}
\label{fig1}
\end{figure}

In addition to the discussed synaptic stochasticity, that we represent here
by means of the variable $x,$ there are independent causes for assuming an
stochastic dynamics of the neuron system. That is, a neuron may sometimes
remain silent even if it endures a large current. This is naturally modelled
by introducing a \textquotedblleft temperature\textquotedblright\ parameter $%
T.$ In practice, one usually assigns a probability which depends on $\left(
h_{i}-\theta _{i}\right) /T,$ where $\theta _{i}$ is a threshold, to the
change according to $\mathrm{sig}(h_{i})=s_{i}$ at time $t.$ This mechanism
is equivalent to assume the existence of a hypothetical \textquotedblleft
thermal bath\textquotedblright\ which induces stochasticity of the neuron
activity by means of a master equation. In general, this equation implies a
tendency towards equilibrium. However, in the present case, the canonical
tendency competes with the stochastic changes of $h_{i},$ which impedes
equilibrium, and the system goes asymptotically to a non-equilibrium steady
state \cite{marroB}. This complex, non-equilibrium situation is at the
origin of the intriguing behavior we describe next.

\section{Computer simulations}

The above programme was implemented in the computer by iterating the
following Monte Carlo algorithm:

\begin{figure}[tbp]
\centerline{
\psfig{file=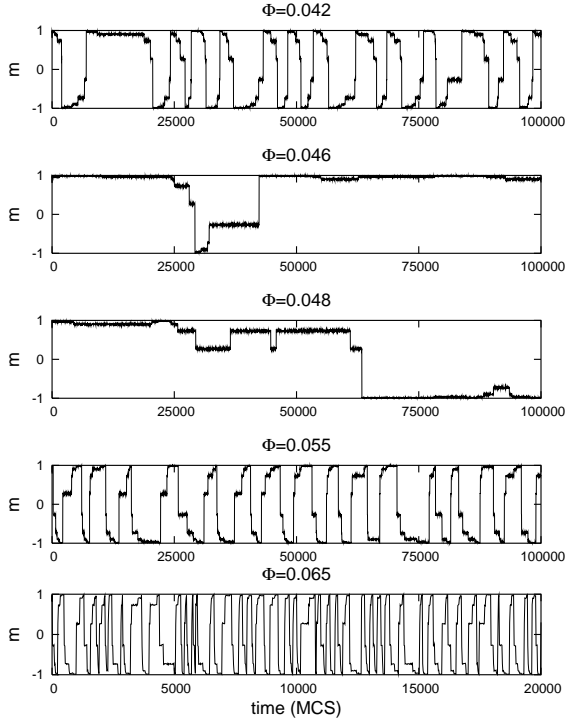,width=7.5cm}
}
\caption{ {\protect\small Monte Carlo simulations in the absence of thermal
fluctuations, $T=0,$ for a single stored pattern and $N=3600$ neurons
showing the effect of varying the synaptic noise parameter $\Phi .$ The
resulting hopping shows dramatic variations of temporal scale and degree of
complexity as one varies $\Phi .$}}
\label{fig2}
\end{figure}

\begin{enumerate}
\item Store $M$ different patterns $\xi _{i}^{\mu }$ in the average weights $%
w_{ij}^{\mathrm{L}}$ according to the chosen, e.g., Hebb's \textit{learning
rule}.

\item Set any state $\mathbf{S}=\left\{ s_{i}\right\} $ at random.

\item Compute the $N$ local fields $\overline{h_{i}}(\mathbf{S})$ as defined
in (\ref{lfaprox}).

\item Choose a site (neuron) at random, repeat the choice $N$ times and keep
only the $n<N$ sites which differ from each other (this procedure lets you
with $n\approx \frac{2}{3}N$ sites ---for the values of $N$ of interest
here).\footnote{%
Both Monte Carlo simulations and analytical results \cite{cortesELSEWHERE}
are in full-agreement and, in the thermodynamic limit, one has that $%
n/N=1-1/e$.}

\item Perform the changes $s_{i}\rightarrow -s_{i}$ at the chosen $n$ sites
using the standard rate $\omega (s_{i}^{\prime }\rightarrow s_{i})=\frac{1}{2%
}\left\{ 1-s_{i}^{\prime }\tanh \left[ \beta \overline{h_{i}}(\mathbf{S}%
^{\prime })\right] \right\} .$

\item Increase time in one unit, and go to step $3$.
\end{enumerate}

\begin{figure}[tbp]
\centerline{
\psfig{file=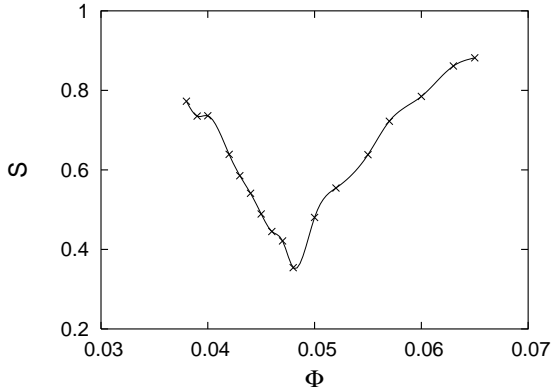,width=7.5cm}
}
\caption{{\protect\small The entropy function, as defined in the main text,
for different time series obtained during Monte Carlo simulations of neural
automata for different values of the synaptic noise parameter $\Phi $ .
Decreasing values of the entropy indicate a tendency towards regularization
of the complexity of the time series. The graph reveals different regimes of
chaoticity.}}
\label{fig3}
\end{figure}

Figure \ref{fig1} illustrates the resulting behavior for a single pattern,
i.e., it corresponds to the limit $\alpha \rightarrow 0.$ This shows a
complex hopping process between the pattern, $\mathbf{\xi ^{1}}$, and the 
\textit{anti-pattern}, $-\mathbf{\xi ^{1}.}$ The figure compares the
evolution at a finite temperature with that in the absence of thermal
fluctuations to demonstrate that hopping is not a consequence of the latter.
Consequently, in order to avoid the short--length oscillations shown in the
bottom graph of figure \ref{fig1}, which are induced by the thermal noise,
we are concerned in the following with simulations at $T=0$.

Figure \ref{fig2} illustrates a main result, namely, that the frequency and
other details of the hopping strongly depend on the value of the parameter $%
\Phi $ which modulates the fast synaptic noise. An appropriate measure of
the associated entropy will provide a quantitative description of the
complexity of this hopping. Using standard fast Fourier transform
algorithms, we computed the power spectra $P(\eta ).$ The normalized
probability $p_{\eta }=P(\eta )/\mathop{\textstyle \sum }_{\eta }P(\eta )$
then allows one to define a regular entropy as $S\equiv -\sum_{\eta }p_{\eta
}\log _{2}p_{\eta }$. This quantity has been used before to detect
regularity out of chaotic activity in actual neurons \cite{torresENTROPY}.
As a matter of fact, $S>0$ is to be associated with chaotic behavior while $%
S=0$ would correspond to periodic dynamics.

Figure \ref{fig3} depicts the entropy which results in our case as a
function of $\Phi $. This shows a minimum which corresponds to the smallest
degeneration in the time series of figure \ref{fig2} (second graph from the
top). Decreasing $S$ indicates a tendency to regularization or smaller
chaoticity, while higher chaos and irregularity in the time series
corresponds to larger values of $S.$

\section{Conclusions}

We have introduced a class of \textit{hybrid} neural automata with two main
features. On one hand, these models provide a convenient arena to analyze
the influence of fast synaptic noise on the retrieval process. On the other
hand, they may describe a continuous transition from sequential,
single--neuron updating to the case of Little dynamics or parallel updating
as one varies the model parameter $n.$ The synaptic noise is modelled trying
to mimic recent observations, namely, the noise occurs in a short--time
scale and conveniently couples to the neuron activity to induce synaptic 
\textit{depression}. Depending on the intensity of this depression, the
model exhibits a varied emergent behavior, including chaotic hopping between
the attractors. This results in a rather complex pattern of neural activity.
Monitoring the entropy suggests how a fast noise might provide a mechanism
to control chaos in living systems \cite{garfinkelSCIENCE}, \cite%
{schiffNATURE}. The design of a mechanism in which noise intensity varies
autonomously could be useful to the self-control of chaos. Notice in this
respect that manipulating $n$ in the model might be convenient for the
purpose. That is, two main cases follow altogether from the present analysis
and some previous work \cite{cortesNC}, \cite{marroPRE}. (1) $n=1,$ for
which the system is sensible to an external stimulus, which may destabilize
the attractor, but it does not exhibit autonomous hopping between
attractors; and (2) $n>1,$ for which hopping occurs autonomously, without
the need for any external stimulus. In the latter case, as far as $n<N,$ the
parameter $\Phi $ allows for a control of the hopping, while this always
occurs at high frequency for $n=N.$ For $n\approx \frac{2}{3}N,$ the case
for which we report some results here, the time the neuron activity stays at
or near each attractor may be varied by tuning $\Phi ,$ as illustrated in
figure \ref{fig2}.

We acknowledge financial support from MEyC and FEDER (project FIS2005-00791
and a \textit{Ram\'{o}n y Cajal} contract).


\end{document}